\documentclass[pra,twocolumn,showpacs,preprintnumbers,amsmath,amssymb]{revtex4}
\usepackage{amssymb}
\usepackage{amsfonts}
\usepackage{tipa}

\usepackage{epsfig,graphicx}
\usepackage{amstext}
\usepackage{amsmath}
\usepackage{graphicx}

\begin{document}


\title{Disentanglement, Bell-nonlocality violation and teleportation capacity of the decaying tripartite states}

\author{Ming-Liang Hu}
\email{mingliang0301@163.com}
\address{School of Science, Xi'an University of Posts and
         Telecommunications, Xi'an 710061, China}

\begin{abstract}
 Dynamics of disentanglement as measured by the tripartite negativity and Bell
 nonlocality as measured by the extent of violation of the multipartite Bell-type
 inequalities are investigated in this work. It is shown definitively that for the
 initial three-qubit Greenberger-Horne-Zeilinger (GHZ) or {\it W} class state preparation
 the Bell nonlocality suffers sudden death under the influence of thermal reservoirs.
 Moreover, all the Bell-nonlocal states are useful for nonclassical teleportation,
 while there are entangled states that do not violate any Bell-type inequalities, but
 still yield nonclassical teleportation fidelity.
\end{abstract}

\pacs{03.65.Ud, 03.67.-a, 03.65.Ta \\Key Words: Disentanglement;
Bell inequality; Teleportation}

\maketitle

\section{Introduction}
Entanglement and Bell nonlocality are two basic ingredients of a
quantum state which are intimately related to each other
\cite{Nielsen}. Their relations have been the research interests for
many years and remain still an important subject deserving to be
investigated. Entanglement, which describes correlations between two
or more subsystems, refers to the state of a composite system that
cannot be written as products of states of each subsystem. As a
physical resource, entanglement is crucial for nearly all
applications related to quantum information processing (QIP). One
such application is quantum teleportation \cite{Bennett}, by which
an unknown state can be transmitted from the sender to a distant
receiver with the help of local operations and classical
communication (LOCC). However, not all the states that are entangled
can be used for teleportation with fidelity (see sections below)
better than that achievable via classical communication alone, and
the fidelity is even not a monotonic function of the degree of
entanglement of the resource \cite{Popescu,Horodecki,Yeo}. This
demonstrates that entanglement may only reveals certain aspects of a
quantum state.

Bell nonlocality corresponds to another quantum correlation of a
quantum state that cannot be reproduced by any classical local
hidden variable models. This nonlocal property are manifested
unambiguously by violation of different Bell-type inequalities,
which plays a fundamental role in better understanding of the subtle
aspects of quantum mechanics \cite{Nielsen,Clauser}. Historically,
the violation of Bell inequalities has been considered as a means of
determining whether there is entanglement between two qubits, for
the inseparability of a bipartite pure state corresponds to the
violation of the Bell inequality in the Clauser-Horne-Shimony-Holt
(CHSH) form, and vice versa \cite{Clauser}. But this is not the case
for the mixed states (which are in practice the ones always
encountered). As demonstrated initially by Werner \cite{Werner1},
there exist bipartite mixed states which are entangled but do not
violate any Bell-type inequalities. Further studies also showed that
the maximal violation of a Bell inequality does not behave
monotonously under LOCC \cite{Popescu,Gisin1}.

The ability for transmitting information reveals also an important
aspect of nonseparability of a quantum state which is intimately
related to entanglement and Bell-nonlocal correlations
\cite{Popescu}. For bipartite two-qubit systems, it has been
demonstrated that all states that violate the CHSH form of Bell
inequality can be used for teleporting an arbitrary one-qubit state
with nonclassical fidelity \cite{Horodecki}, while there are
entangled mixed states which do not violate any Bell-type
inequalities, but still yield nonclassical teleportation fidelity
\cite{Popescu}. For the tripartite systems, as we know, there are
two distinct classes of multipartite entangled states, i.e., the
Greenberger-Horne-Zeilinger (GHZ) class and the {\it W} class, which
bear incompatible multipartite correlations in the sense that they
cannot be transformed into each other under stochastic local
operations and classical communication \cite{Dur}. Particularly, it
has been demonstrated that besides the Bell states, the three-qubit
GHZ and {\it W} class states in the form of $|\psi_{\rm
GHZ}\rangle=(|000\rangle+|111\rangle)/\sqrt{2}$ and
$|\psi_{W}\rangle=(\sqrt{2}|001\rangle+|010\rangle+|100\rangle)/2$
can also be adopted as quantum channels for perfect teleportation
under ideal circumstance \cite{Karlsson,Agrawal}. Since they both
are unavoidable disturbed by the surrounding environments, it is
natural to ask for their robustness against decoherence in terms of
their teleportation capacity, and its possible relations with the
degrees of entanglement and Bell-nonlocality violation.

From an applicative point of view, one may hopes that quantum
correlations which are crucial for QIP can be maintained for
sufficiently long times to permit designed tasks to be fulfilled.
But in practice, every quantum system is open and susceptible to the
unavoidable interaction with its surroundings \cite{Breuer,Yu,Jung}.
This may leads to decoherence and destruction of correlations.
Particularly, under certain circumstances, the entanglement of a
bipartite state can even terminate abruptly in a finite time, a
phenomenon termed entanglement sudden death (ESD) by Yu and Eberly
\cite{Yu} and has been recently confirmed experimentally
\cite{Almeida}. Moreover, due to its practical applications in QIP
\cite{Acin,Gisin2}, the research interest in the decay dynamics of
Bell-type correlations has been renewed. Analogous to ESD, it has
also been demonstrated that an initial nonlocal quantum state may
loses its nonlocal property in a finite timescale under the
influence of external forces. This phenomenon has been named
Bell-nonlocality sudden death (BNSD) \cite{Jaeger,Ann} and many
theoretical efforts have been devoted to it with different contexts
\cite{Kofman,Bellomo1,Lijq,Liubq1,Liubq2,Altintas}. For instance,
Ann and Jaeger have demonstrated the occurrence of BNSD for the
initial three-qubit {\it W} class state as measured by its violation
of the MABK inequality \cite{Jaeger}, as well as for the initial
generic class of tripartite state as measured by its violation of
the Svetlichny and WWZB inequalities \cite{Ann}.

\section{Theoretical framework}
In this paper, we would like to investigate the phenomena of
disentanglement and BNSD for tripartite states, and their relation
with fidelity of quantum teleportation. Our system consists of three
qubits (here labeled as $K=A,B,C$), each embedded in a thermal
reservoir. To focus exclusively on the decay of entanglement and
Bell nonlocality as they arise from the influence of thermal noise,
we assume the qubits are separated by spatial distances large enough
and thus there are no direct interactions between them, that is,
every qubit interacts only, and independently with its own
environment. Then under the condition of Markovian approximation,
the reduced dynamics of the system state $\rho$ can be described by
a general master equation of Lindblad form \cite{Breuer}
\begin{eqnarray}
 \frac{d\rho}{dt}&=&\frac{1}{2}\sum_{K,m}\gamma_K (2\mathcal {L}_{K,m}\rho\mathcal {L}_{K,m}^\dag
                  \nonumber\\&&
                  -\mathcal {L}_{K,m}^\dag\mathcal {L}_{K,m}\rho
                  -\rho\mathcal {L}_{K,m}^\dag\mathcal {L}_{K,m}),
\end{eqnarray}
where $\gamma_K$ $(K=A,B,C)$ are the damping rates of the qubits due
to their coupling to the reservoir. The generators of decoherence
are given by $\mathcal {L}_{K,1}=\sqrt{\bar{n}+1}\sigma_K^{-}$ and
$\mathcal {L}_{K,2}=\sqrt{\bar{n}}\sigma_K^{+}$, with
$\sigma_K^{\pm}$ being the raising and lowering operators. $\mathcal
{L}_{K,1}$ and $\mathcal {L}_{K,2}$ describe, respectively, decay
and excitation processes, with rates which depend on the
temperature, here parameterized by the average thermal photons
$\bar{n}$ in the reservoir. For the limiting case of vanishing
temperature (i.e., $\bar{n}=0$), only the spontaneous decay term
survives, leading to a purely dissipative process which drives all
initial states to an unique asymptotic pure state, in which the
three qubits are in their ground states. For the opposite case of
infinite temperature (i.e., $\bar{n}\rightarrow \infty$), decay and
excitation occur at exactly the same rate, and the noise induced by
the transitions between the two levels brings the system into a
stationary, maximally mixed state. Physically, the above models can
describe, e.g., three two-level atoms with interatomic separations
larger than the spatial correlation length of the reservoir such
that the collective damping and the collective shift of the atomic
levels are negligible \cite{Breuer}.

Solutions of the above master equation with arbitrary initial
conditions can be derived exactly in several different ways
\cite{Kraus,Bellomo2}, and here we use the operator-sum
representation \cite{Kraus}. In many situations of physical
interest, this representation allows a transparent analysis of
system dynamics without invoking the explicit forms of the initial
conditions. In the standard basis expanded by the eigenstates of the
product Pauli spin operator
$\sigma_A^z\otimes\sigma_B^z\otimes\sigma_C^z$, the reduced density
matrix for the three qubits together is given by the following
completely positive and trace preserving (CPTP) map \cite{Kraus}
\begin{equation}
 \rho(t)=\mathcal {L}[\rho(0)]=\sum_{i,j,k=1}^4 G_{ijk}\rho(0)G_{ijk}^\dag,
\end{equation}
where the time-dependent tensor-product superoperator $\mathcal
{L}=\mathcal {L}_A\otimes\mathcal {L}_B\otimes\mathcal {L}_C$
contains 64 terms. The Kraus operator $G_{ijk}$ describes the
interaction of the qubits with the thermal reservoir and satisfies
the CPTP relation $\sum_{ijk}G_{ijk}^\dag G_{ijk}=1$ for all $t$.
Since the thermal noise operates locally on individual subsystems,
$G_{ijk}$ can be expressed in terms of the tensor products of
$E_i^A$, $E_j^B$ and $E_k^C$ as $G_{ijk}=E_i^A\otimes E_j^B\otimes
E_k^C$. Here $E_i^A$, $E_j^B$ and $E_k^C$ are the Kraus operators
describing time evolution of each qubit alone, and individually
satisfy the usual completeness condition for the operator-sum
decomposition of CPTP maps \cite{Kraus}. Their explicit forms are as
follows
\begin{eqnarray}
 &&E_1^K=\sqrt{\frac{\bar{n}+1}{2\bar{n}+1}}\left(\begin{array}{cc}
                                              p_K   &  0 \\
                                              0   & 1 \\
                                              \end{array}\right),\nonumber\\
 &&E_2^K=\sqrt{\frac{\bar{n}}{2\bar{n}+1}}\left(\begin{array}{cc}
                                              1   &  0 \\
                                              0   & p_K \\
                                              \end{array}\right),\nonumber\\
 &&E_3^K=\sqrt{\frac{\bar{n}}{2\bar{n}+1}}\left(\begin{array}{cc}
                                              0   &  \sqrt{1-p_K^2} \\
                                              0   &  0 \\
                                              \end{array}\right),\nonumber\\
 &&E_4^K=\sqrt{\frac{\bar{n}+1}{2\bar{n}+1}}\left(\begin{array}{cc}
                                              0   &  0 \\
                                              \sqrt{1-p_K^2}   &  0 \\
                                              \end{array}\right),
\end{eqnarray}
where the time-dependent factors $p_K$ $(K=A,B,C)$ appearing in the
above equations are given by $p_K=e^{-(2\bar{n}+1)\gamma_K t/2}$.
Note that when $\bar{n}=0$, only $E_1^K$ and $E_4^K$ survive, so
solution (2) has simple analytical form. We will apply solution (2)
to analyze the effects of the thermal reservoir on the decay
dynamics of entanglement and Bell-inequality violation for the
initial GHZ and {\it W} class states. For simplicity, we will take
the noise properties to be the same for the three qubits such that
$\gamma_A=\gamma_B=\gamma_C=\gamma$ and
$p_A=p_B=p_C=p=e^{-(2\bar{n}+1)\gamma t/2}$.

After obtaining explicit forms of the reduced density matrix
$\rho(t)$, we can discuss dynamics of disentanglement,
Bell-nonlocality violation and the ability of $\rho(t)$ for quantum
teleportation. To describe the disentanglement process, we need a
concrete measure of entanglement contained in a quantum state. The
tripartite negativity $N$, which was introduced by Sab\'{\i}n and
Garc\'{\i}a-Alcaine \cite{Sabin} is particularly convenient for the
case of current interest. It can be calculated explicitly from the
density matrix $\rho(t)$ as
\begin{equation}
 N=(N_{A-BC}N_{B-CA}N_{C-AB})^{1/3},
\end{equation}
where $N_{A-BC}=-\sum_i \mu_i^A$, $N_{B-CA}=-\sum_i \mu_i^B$ and
$N_{C-AB}=-\sum_i \mu_i^C$ are the negativities introduced by Vidal
and Werner \cite{Vidal}, and the sums are taken over all the
negative eigenvalues $\mu_i^K$ of the partial transpose
$\rho^{T_K}(t)$ of $\rho(t)$ with respect to the subsystem $K$. Note
that for the mixed states, $N$ is not able to quantify multipartite
entanglement fully, but its positivity ensures that the state under
consideration is not separable \cite{Sabin}. Thus if the Bell
nonlocality of a quantum state dies out before that of the
tripartite negativity, one can say that it dies out before that of
entanglement.

Moreover, we will consider Bell-type nonlocality in different
contexts and we use two classes of multipartite Bell-type
inequalities to detect the existence of nonlocal correlations as
measured by the extent of their violations. The first one we are
interested in is the Svetlichny inequality $|\langle \mathcal
{S}\rangle_{\rho(t)}|\leqslant 4$ \cite{Svetlichny}, which
distinguishes genuinely tripartite Bell nonlocality associated with
$\rho(t)$. Here $\langle \mathcal {S}\rangle_{\rho(t)}={\rm
tr}[\mathcal {S}\rho(t)]$ is the expectation value of the Svetlichny
operator given by
\begin{eqnarray}
 \mathcal {S}&=&M_A M_B M_C+M_A M_B M'_C+M_A M'_B M_C
             \nonumber\\&&
             +M'_A M_B M_C-M'_A M'_B M'_C-M'_A M'_B M_C
             \nonumber\\&&
             -M'_A M_B M'_C-M_A M'_B M'_C,
\end{eqnarray}
where the measurement operators $M_K$ and $M'_K$ correspond to the
measurements on each of the subsystems $K$, while the primed and
unprimed terms denote the two different directions in which the
corresponding party measures (the same applies also to the WWZB
operators). A quantum state $\rho(t)$ violates the Svetlichny
inequality whenever $|\langle \mathcal {S}\rangle_{\rho(t)}|>4$, and
in quantum mechanics the Svetlichny inequality is violated up to
$|\langle \mathcal {S}\rangle_{\rho(t)}|=4\sqrt{2}$, which is
achieved only when the system is prepared in the maximally entangled
GHZ state \cite{Dur}.

The second inequality convenient for our purpose is the WWZB
Bell-type inequalities \cite{Werner2,Zukowski}. For the three-qubit
system, there are 256-element set of such inequalities, which belong
to five distinct classes due to some basic symmetries, and the
behavior of a single class is identical to that of all members of
that class, i.e., the violation of even a single element of each
class is sufficient for Bell nonlocality of that class
\cite{Werner2}. Thus we only need to consider one inequality from
each of the five distinct classes:
\begin{eqnarray}
 &&\mathcal {B}_{\rm P1}=2M_A M_B M_C,\nonumber\\
 &&\mathcal {B}_{\rm P2}=\frac{1}{2}(-M_A M_B M_C+M_A M_B M'_C+M_A M'_B M_C
                         \nonumber\\&&
                         ~~~~~~~~+M'_A M_B M_C+M_A M'_B M'_C+M'_A M_B M'_C
                         \nonumber\\&&
                         ~~~~~~~~+M'_A M'_B M_C+M'_A M'_B M'_C),\nonumber\\
 &&\mathcal {B}_{\rm P3}=[M_A(M_B+M'_B)+M'_A(M_B-M'_B)]M_C,\nonumber\\
 &&\mathcal {B}_{\rm P4}=M_A M_B(M_C+M'_C)-M'_A M'_B(M_C-M'_C),\nonumber\\
 &&\mathcal {B}_{\rm P5}=M_A M_B M'_C+M_A M'_B M_C+M'_A M_B M_C
                         \nonumber\\&&
                         ~~~~~~~~\,-M'_A M'_B M'_C.
\end{eqnarray}

For nonlocal quantum states, there should be at least one of the
$|\langle\mathcal {B}_{\rm PI}\rangle_{\rho(t)}|>2$ $({\rm
I}=1,2,3,4,5)$, where $\langle\mathcal {B}_{\rm
PI}\rangle_{\rho(t)}={\rm tr}[\mathcal {B}_{\rm PI}\rho(t)]$. For
the behavior of a system to be describable by a fully local hidden
variable model, however, all of the WWZB set of inequalities must be
satisfied jointly.

To characterize the quality of the teleported output state
$\rho_{\rm out}$ under the influence of thermal reservoir, we
calculate the average fidelity (the fidelity
$F(\theta,\phi)=\langle\varphi_{\rm in}|\rho_{\rm out}|\varphi_{\rm
in}\rangle$ averaged over all pure input states on the Bloch sphere)
\cite{Nielsen}, defined as
\begin{equation}
F_{\rm av}={1\over 4\pi}
   \int_0^{2\pi}d\phi \int_0^{\pi}d\theta \sin\theta F(\theta,\phi),
\end{equation}
where $|\varphi_{\rm
in}\rangle=\cos(\theta/2)|0\rangle+e^{i\phi}\sin(\theta/2)|1\rangle$
is the input state needs to be teleported, with $0\leqslant
\theta\leqslant \pi$ and $0\leqslant \phi\leqslant 2\pi$ being the
polar and azimuthal angles, respectively. When the aforementioned
initial GHZ class state $|\psi_{\rm
GHZ}\rangle=(|000\rangle+|111\rangle)/\sqrt{2}$ or the initial {\it
W} class state
$|\psi_W\rangle=(\sqrt{2}|001\rangle+|010\rangle+|100\rangle)/2$ is
used as quantum channel, by following the methodology of Ref.
\cite{Jung}, the average teleportation fidelity at an arbitrary time
$t$ can be derived readily as $F_{\rm av}(\rho_{\rm
GHZ})=(1+\rho_{\rm GHZ}^{11+44+55+88}+2\rho_{\rm GHZ}^{18})/3$ and
$F_{\rm
av}(\rho_W)=(1+\rho_{W}^{22+33+44+88+35-46}+2\sqrt{2}\rho_{W}^{23})/3$,
where the abbreviations $\rho_{\rm \Pi}^{ij\pm kl\pm
mn\pm\ldots}=\rho_{\rm \Pi}^{ij}(t)\pm \rho_{\rm \Pi}^{kl}(t)\pm
\rho_{\rm \Pi}^{mn}(t)\pm\ldots$, with $\rho_{\rm \Pi}^{ij}(t)$
($\Pi={\rm GHZ}$ or $W$) being the elements of the density matrix
$\rho_{\rm \Pi}(t)$. Since the noise imposed by the thermal
reservoir gives rise both to decoherence and to entanglement losses,
unit fidelity cannot be achieved for this case.

\section{Entanglement, Bell-inequality violation and teleportation dynamics}
Now we begin our discussion about disentanglement dynamics and
teleportation capacity for the system prepared initially in the GHZ
class state $|\psi_{\rm GHZ}\rangle$. The density matrix at an
arbitrary time $t$ can be obtained directly from Eqs. (2) and (3),
and the nonvanishing elements are $\rho_{\rm GHZ}^{11-88}(t)$ (i.e.,
the diagonal elements) and $\rho_{\rm GHZ}^{18,81}(t)$. Combination
of this with Eqs. (4) and (7), one can derive the tripartite
negativity $N(\rho_{\rm GHZ})$ and the average fidelity $F_{\rm
av}(\rho_{\rm GHZ})$ analytically as
\begin{eqnarray}
 &&N(\rho_{\rm GHZ})=\frac{1}{2}{\rm max}\{0,~\sqrt{\alpha^2+p^6}-\beta\},\nonumber\\
 &&F_{\rm av}(\rho_{\rm GHZ})=\frac{1}{2}+\frac{p^3}{3}+\frac{p^4}{6}+\frac{(1-p^2)^2}{6(2\bar{n}+1)^2},
\end{eqnarray}
where the corresponding parameters $\alpha$ and $\beta$ appeared in
the above equations are given by
\begin{eqnarray}
 &&\alpha=\frac{(1-p^2)[2\bar{n}(\bar{n}+1)(3p^4-1)+2p^4-p^2]}{2(2\bar{n}+1)^3},\nonumber\\
 &&\beta=\frac{(1-p^2)[2\bar{n}(\bar{n}+1)(p^2+1)+p^2]}{2(2\bar{n}+1)^2},
\end{eqnarray}
with the factor $p$ being defined below Eq. (3).
\begin{figure}
\centering
\resizebox{0.45\textwidth}{!}{%
\includegraphics{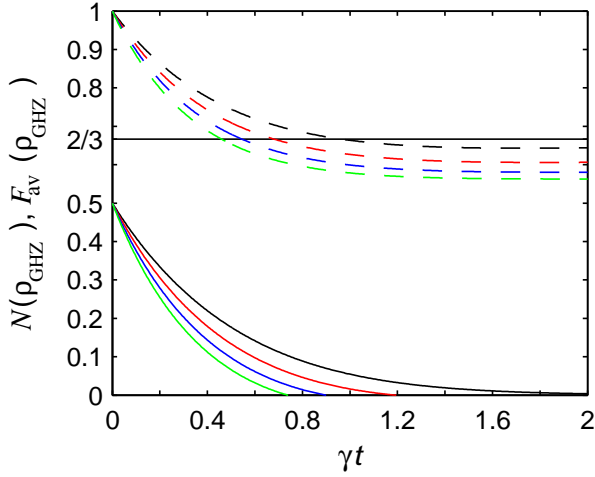}}
\caption{(Color online) Tripartite negativity $N(\rho_{\rm GHZ})$
(solid curves) and average fidelity $F_{\rm av}(\rho_{\rm GHZ})$
(dashed curves) versus $\gamma t$. For every line style, the curves
from top to bottom correspond to the cases of $\bar{n}=0$,
$\bar{n}=0.1$, $\bar{n}=0.2$ and $\bar{n}=0.3$.} \label{fig:1}
\end{figure}

Plots of the tripartite negativity $N(\rho_{\rm GHZ})$ and average
fidelity $F_{\rm av}(\rho_{\rm GHZ})$ versus the rescaled time
$\gamma t$ are displayed in Fig. 1 with different $\bar{n}$. One can
observe that the evolution of $N(\rho_{\rm GHZ})$ shows an
exponentially decaying behavior, and it is very sensitive to the
variations of the reservoir temperature. From Eq. (8) one can see
that the entanglement measured by the tripartite negativity
disappears if $\alpha^2+p^6\leqslant \beta^2$. When $\bar{n}=0$,
this simplifies to $1-p^2\geqslant 1$. Since for all finite values
of $\gamma$ the factor $p$ approaches zero exponentially only in the
infinite time limit, the sudden death of $N(\rho_{\rm GHZ})$ does
not happen for this special case. This behavior is clearly
illustrated by the black solid curve shown in Fig. 1. When
$\bar{n}\neq0$, however, $N(\rho_{\rm GHZ})$ behaves very
differently. As can be seen from Fig. 1, it ceases to exist in a
finite timescale $\tau_{\rm E}$ which is shortened gradually by
increasing the values of $\bar{n}$. This implies that the
devastating effects of the thermal reservoir on entanglement of the
system becomes severe and severe with increasing temperature.
Furthermore, it should be note that for the initial GHZ state
$|\psi_{\rm GHZ}\rangle$, $N(\rho_{\rm
GHZ})=N_{A-BC}=N_{B-CA}=N_{C-AB}$, thus the sudden death of
$N(\rho_{\rm GHZ})$ indicates the sudden death of $N_{A-BC}$,
$N_{B-CA}$ and $N_{C-AB}$.
\begin{figure}
\centering
\resizebox{0.45\textwidth}{!}{%
\includegraphics{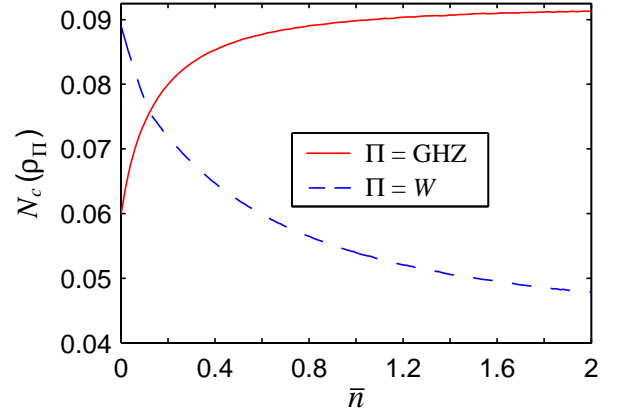}}
\caption{(Color online) Critical tripartite negativities
$N_c(\rho_{\rm GHZ})$ (solid curves) and $N_c(\rho_{W})$ (dashed
curves) versus $\bar{n}$.} \label{fig:2}
\end{figure}

When considering robustness of the initial GHZ state as a quantum
channel for teleportation, as can be seen from the dashed curves
shown in Fig. 1, $F_{\rm av}(\rho_{\rm GHZ})$ decays to the
classical limiting value of $2/3$ after a critical time $\tau_{\rm
T}$, which decreases with increasing $\bar{n}$, and when $\bar{n}=0$
one can obtain $\gamma\tau_{\rm T}=\ln[(3+\sqrt{5})/2]$. Moreover,
one can note that the critical time $\tau_{\rm T}$ is earlier than
the death time $\tau_{\rm E}$ of tripartite negativity for any fixed
$\bar{n}$. Since a non-zero tripartite negativity signals the
entanglement of the system, this phenomenon indicates that not all
the three-qubit entangled states generated from the initial GHZ
class state are useful for nonclassical teleportation. In fact, a
minimum tripartite negativity (arrives at the critical time
$\tau_{\rm T}$) is always necessary for the achievement of
nonclassical fidelity for the situations considered here. As
displayed evidently in Fig. 2, the critical tripartite negativity
$N_c(\rho_{\rm GHZ})$ after which the teleportation protocol fails
to achieve a nonclassical fidelity is increased by increasing the
reservoir temperature. This phenomenon is mainly caused by the
competition between the increased temperature $\bar{n}$ and the
decreased critical time $\tau_{\rm T}$. Because the increase of
$\bar{n}$ always decrease $\tau_{\rm T}$ and $N(\rho_{\rm GHZ})$,
but meanwhile the decrease of $\tau_{\rm T}$ always increase
$N(\rho_{\rm GHZ})$. Due to the complexity of Eqs. (8) and (9), it
is difficult to obtain an analytic form of $N_c (\rho_{\rm GHZ})$,
however, we can make a heuristic analysis for two limiting cases,
i.e., $\bar{n}=0$ and $\bar{n}\rightarrow \infty$. For $\bar{n}=0$,
since $\gamma\tau_{\rm T}=\ln[(3+\sqrt{5})/2]$, we obtain $N_c
(\rho_{\rm GHZ})=(2-\sqrt{5}+\sqrt{197-88\sqrt{5}})/4\simeq 0.0598$,
while for $\bar{n}\rightarrow \infty$, $\gamma\tau_{\rm T}$ can only
be solved numerically from the nonlinear equation $p^4+2p^3=1$,
which gives rise to $p\simeq 0.7166$ and $N_c (\rho_{\rm GHZ})
\simeq 0.0920$. This analysis corroborates our finding presented in
Fig. 2.
\begin{figure}
\centering
\resizebox{0.45\textwidth}{!}{%
\includegraphics{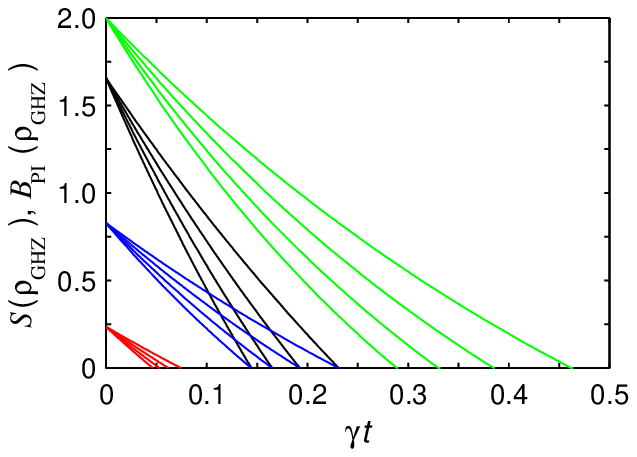}}
\caption{(Color online) Bell-nonlocality violation of $\rho_{\rm
GHZ}(t)$. Here the black, red, blue and green curves show
respectively dynamics of $S(\rho_{\rm GHZ})$, $B_{\rm P2}(\rho_{\rm
GHZ})$, $B_{\rm P3}(\rho_{\rm GHZ})$ and $B_{\rm P5}(\rho_{\rm
GHZ})$. For every line color, the curves from right to left
correspond to the cases of $\bar{n}=0$, $\bar{n}=0.1$, $\bar{n}=0.2$
and $\bar{n}=0.3$.} \label{fig:3}
\end{figure}

As mentioned before, Bell-inequality violations may act as an
indicator of the usefulness of entanglement. Let us discuss now this
issue by quantum states evolving in time. Our aim is to individuate
the Svetlichny inequality violation regions characterized by
$|\langle \mathcal {S}\rangle_{\rho_{\rm GHZ}(t)}|>4$ and the WWZB
inequality violation regions characterized by $|\langle \mathcal
{B}_{\rm PI}\rangle_{\rho_{\rm GHZ}(t)}|>2$ $({\rm I}=1,2,3,4,5)$,
and compare their possible relations with disentanglement and
average teleportation fidelity. For this purpose, we calculate the
expectation values of the Svetlichny operator $\mathcal {S}$ and the
WWZB operators $ \mathcal {B}_{\rm PI}$. For the initial GHZ state,
the measurement operators for the first qubit (i.e., qubit {\it A})
are defined as $M_A\equiv\sigma_y$ and $M'_A\equiv\sigma_x$
\cite{Ann}, while the measurement operators for the second and the
third qubits are defined with respect to the first one by a rotation
\begin{eqnarray}
 \left(\begin{array}{cc}
        M_K \\
        M'_K \\
       \end{array}
 \right)
 =
 \left(\begin{array}{cc}
        \cos\theta_K  &  -\sin\theta_K \\
        \sin\theta_K  &   \cos\theta_K \\
       \end{array}
 \right)
 \left(\begin{array}{cc}
        M_A \\
        M'_A \\
       \end{array}
 \right),
\end{eqnarray}
where $\theta_K$ $(K=B,C)$ are the rotation angles. Combination of
this with Eq. (5) one can obtain
\begin{equation}
 \langle \mathcal {S}\rangle_{\rho_{\rm GHZ}(t)}=4p^3 (\sin\theta_{BC}-\cos\theta_{BC}),
\end{equation}
where $\theta_{BC}=\theta_{B}+\theta_{C}$, and the same notation
will be used throughout this paper. Recall that if $|\langle
\mathcal {S}\rangle_{\rho_{\rm GHZ}(t)}|>4$, the state $\rho_{\rm
GHZ}(t)$ is genuinely tripartite Bell nonlocal \cite{Ann}. Since
$p=e^{-(2\bar{n}+1)\gamma t/2}$ and the modulus of the trigonometric
term takes its maximum $\sqrt{2}$ whenever $\theta_{BC}=-\pi/4$ or
$3\pi/4$, the maximum expectation value of the Svetlichny operator
$\mathcal {S}$ evolves according to $|\langle \mathcal
{S}\rangle_{\rho_{\rm GHZ}(t)}|=4\sqrt{2}e^{-3(2\bar{n}+1)\gamma
t/2}$, and approaches the classical threshold value 4 in a finite
timescale $\tau_{\rm S}=\ln 2/[3(2\bar{n}+1)\gamma]$. As can be seen
from the dynamics of $S(\rho_{\rm GHZ})={\rm max}\{|\langle \mathcal
{S}\rangle_{\rho_{\rm GHZ}(t)}|-4,0\}$ presented in Fig. 3, the
temperature of the thermal reservoir can influence $S(\rho_{\rm
GHZ})$ to a great extent, and the time regions for genuinely
tripartite nonlocality shrink with increasing temperature.
Particularly, in the small $\gamma t$ region we find $|\langle
\mathcal {S}\rangle_{\rho_{\rm GHZ}(t)}|$ decays up to quadratic
terms in time as $|\langle \mathcal {S}\rangle_{\rho_{\rm
GHZ}(t)}|=2\sqrt{2}(a^2-2a+2)+\mathcal {O}(t^3)$, where
$a=3(2\bar{n}+1)\gamma t/2$. The relative difference between the
actual and the approximate results, is in practice negligible when
both are above the classical threshold $|\langle \mathcal
{S}\rangle_{\rho_{\rm GHZ}(\tau_{\rm S})}|=4$.

Next we extend the analysis of Bell-type nonlocal correlations in
tripartite states addressed by violation of the WWZB inequalities.
The expectation values for the five distinct classes of the
$\mathcal {B}_{\rm PI}$ operators for the state $\rho_{\rm GHZ}(t)$
can also be calculated analytically, which are given by
\begin{eqnarray}
 &&\langle\mathcal {B}_{\rm P1}\rangle_{\rho_{\rm GHZ}(t)}=\langle\mathcal {B}_{\rm P4}\rangle_{\rho_{\rm GHZ}(t)}
                                                            =2p^3\sin\theta_{BC},\nonumber\\
 &&\langle\mathcal {B}_{\rm P2}\rangle_{\rho_{\rm GHZ}(t)}=-p^3 (2\sin\theta_{BC}+\cos\theta_{BC}),\nonumber\\
 &&\langle\mathcal {B}_{\rm P3}\rangle_{\rho_{\rm GHZ}(t)}=2p^3 (\sin\theta_{BC}-\cos\theta_{BC}),\nonumber\\
 &&\langle\mathcal {B}_{\rm P5}\rangle_{\rho_{\rm GHZ}(t)}=-4p^3\cos\theta_{BC}.
\end{eqnarray}

Recalling that $p=e^{-(2\bar{n}+1)\gamma t/2}$, one sees immediately
that the state $\rho_{\rm GHZ}(t)$ does not violate the WWZB-type
inequalities with respect to the operators $\mathcal {B}_{\rm P1}$
and $\mathcal {B}_{\rm P4}$ in the full time region. For the
remaining three classes, plots of $B_{\rm PI}(\rho_{\rm GHZ})={\rm
max}\{|\langle\mathcal {B}_{\rm PI}\rangle_{\rho_{\rm
GHZ}(t)}|-2,0\}$ $({\rm I}=2,3,5)$ versus the rescaled time $\gamma
t$ are displayed in Fig. 3 as red, blue and green curves with
different $\bar{n}$. Clearly, they show qualitatively similar
dynamical behaviors, the most pronounced difference is the time
regions during which the corresponding inequalities are violated.
For the inequality of form ${\rm P2}$, since the trigonometric term
is strictly bounded by $\sqrt{5}$, the maximum expectation value of
$\mathcal {B}_{\rm P2}$ evolves as $|\langle\mathcal {B}_{\rm
P2}\rangle_{\rho_{\rm GHZ}(t)}|=\sqrt{5}e^{-3(2\bar{n}+1)\gamma
t/2}$, and approaches the critical value 2 in a short timescale
$\tau_{B_{\rm P2}}=\ln(5/4)/[3(2\bar{n}+1)\gamma]$, which decreases
gradually with increasing temperature (cf. the red curves shown in
Fig. 3). Second, for the inequality of form ${\rm P3}$, the
trigonometric term is strictly limited by $\sqrt{2}$, and the
maximum expectation value of $\mathcal {B}_{\rm P3}$ is
$|\langle\mathcal {B}_{\rm P3}\rangle_{\rho_{\rm
GHZ}(t)}|=2\sqrt{2}e^{-3(2\bar{n}+1)\gamma t/2}$, who becomes
smaller than 2 in precisely the same timescale as that of the
Svetlichny inequality, i.e., $\tau_{B_{\rm P3}}=\tau_{\rm S}$ .
Finally, for the inequality of form ${\rm P5}$, the maximum
$|\langle\mathcal {B}_{\rm P5}\rangle_{\rho_{\rm
GHZ}(t)}|=4e^{-3(2\bar{n}+1)\gamma t/2}$ decays to its critical
value 2 in a finite time $\tau_{B_{\rm
P5}}=\ln4/[3(2\bar{n}+1)\gamma]$. Note that after time $\tau_{B_{\rm
P5}}$ all the five sets of the WWZB inequalities are satisfied, and
the state $\rho_{\rm GHZ}(t)$ becomes Bell local, i.e., its
correlations can be reproduced by a classical local model
\cite{Nielsen}. This implies that under the influence of thermal
reservoir, the phenomenon of BNSD occurs for definitive times for
the initial GHZ class state preparation.

Moreover, by comparing Fig. 3 with Fig. 1, one can also note that
the lifetime for the tripartite negativity is longer than the time
region during which the teleportation protocol outperforms those of
classical ones or the lifetime of Bell nonlocality addressed by
violation of the Svetlichny as well as the full set of WWZB
inequalities. Since a non-zero tripartite negativity signals the
fact that the system is entangled, this phenomenon indicates that
for the initial GHZ state preparation, all the Bell-nonlocal (not
necessary the genuinely tripartite Bell nonlocal) states are
entangled and can be used for nonclassical teleportation, while
there are still some tripartite states (i.e., the states locating
between the time region $\tau_{\rm T}$ and $\tau_{\rm E}$) which are
entangled but fail to achieve a nonclassical teleportation fidelity.
This conclusion is very similar to that of the two-qubit case
\cite{Popescu,Horodecki,Yeo}, for which the subset of the entangled
X-type states that violate the Bell-CHSH inequality can always be
used for nonclassical teleportation.
\begin{figure}
\centering
\resizebox{0.45\textwidth}{!}{%
\includegraphics{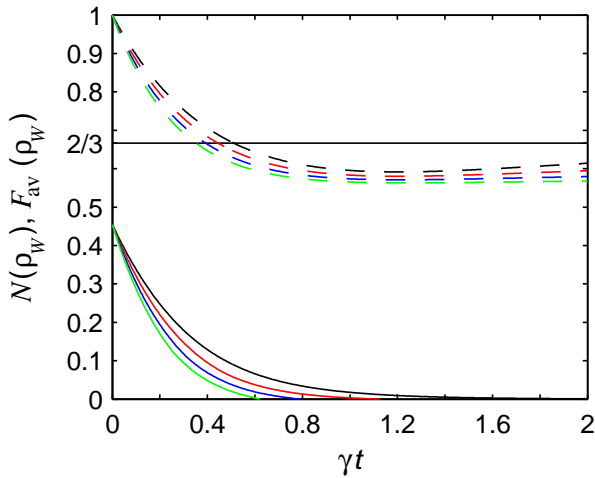}}
\caption{(Color online) Tripartite negativity $N(\rho_W)$ (solid
curves) and average fidelity $F_{\rm av}(\rho_{W})$ (dashed curves)
versus $\gamma t$. For every line style, the curves from top to
bottom correspond to the cases of $\bar{n}=0$, $\bar{n}=0.1$,
$\bar{n}=0.2$ and $\bar{n}=0.3$.} \label{fig:4}
\end{figure}

We now turn our attention to an analysis of disentanglement, Bell
nonlocality and teleportation capacity for the initial {\it W} state
preparation. The density matrix $\rho_W(t)$ can also be obtained
directly from Eqs. (2) and (3), from which the complete analytical
forms of the tripartite negativity and the average fidelity can be
derived exactly, however, we do not list them here explicitly
because their expressions are quite involved. Instead, we plot
numerically in Fig. 4 the time behaviors of both $N(\rho_W)$ and
$F_{\rm av}(\rho_{W})$ for the same values of the system parameters.

The solid curves in Fig. 4 show the numerical results for
$N(\rho_W)$ versus $\gamma t$ with different $\bar{n}$, from which
one can note a pronounced difference between the dynamical behaviors
for the zero and nonzero temperature cases. Similar to that for the
initial GHZ state preparation, the tripartite negativity $N(\rho_W)$
decays exponentially and disappears only in the infinite time limit
for the zero temperature reservoir, thus there is no ESD happens for
this special case. For the situations of nonzero temperature
reservoir, however, the tripartite negativity terminates abruptly in
a finite time $\tau_{\rm E}$, which decreases with increasing
temperature and is shorter than that for the initial GHZ state
preparation (cf. Fig. 4 and Fig. 1). Thus one can say that the
tripartite negativity of the initial {\it W} state is fragile
compared with that of the initial GHZ state in the sense that it
decays in a faster rate. But it should be note that this does not
mean the fragility of the entanglement for the initial {\it W}
state, for the disappearance of the tripartite negativity cannot
guarantee that the state is always separable. Moreover, we would
like to mention here that for the special case of infinite
temperature reservoir (i.e., $\bar{n}\rightarrow \infty$), Carvalho
et al. \cite{Carvalho} have shown that the entanglement measured by
the multipartite concurrence for the initial standard {\it W} state
preparation also dies faster than that for the initial GHZ state
preparation.

From the dashed curves shown in Fig. 4 one can see that in the whole
temperature region, the teleportation protocol loses its quantum
advantage over purely classical communication in a finite timescale
$\tau_{\rm T}$, which is significantly shorter than that for the
initial GHZ state preparation, and $F_{\rm av}(\rho_{W})$ is always
smaller than $F_{\rm av}(\rho_{\rm GHZ})$ when $t>0$. Thus although
they both ensure unit fidelity under ideal circumstance
\cite{Karlsson,Agrawal}, the quantum channel with the initial GHZ
class state preparation outperforms that of the initial {\it W}
class state preparation in terms of their teleportation capacity.
Moreover, from Fig. 4 one can also observe that a non-vanishing
tripartite negativity $N_c(\rho_W)$ is always necessary for $F_{\rm
av}(\rho_{W})>2/3$, however, as can be seen from the blue dashed
curve shown in Fig. 2, $N_c(\rho_W)$ decreases exponentially with
increasing $\bar{n}$ and thus behaves in a markedly different way
compared to $N_c(\rho_{\rm GHZ})$, which can also be understood from
the competitive mechanism between the increased temperature
$\bar{n}$ and the decreased critical time $\tau_{\rm T}$. Although
the complexity of $\rho_{W}(t)$ prevents us to derive an analytic
form of $N_c(\rho_W)$ versus $\bar{n}$ for general case, we have
$F_{\rm av}(\rho_{W})=5p^4/6-p^2/2+2/3$ for $\bar{n}=0$, which
yields $\gamma\tau_{\rm T}=\ln(5/3)$ and $N_c(\rho_W)\simeq 0.0891$,
while for very large $\bar{n}$ we have $\bar{n}\gamma\tau_{\rm
T}\simeq0.3257$ and $N_c(\rho_W)\simeq 0.0404$.
\begin{figure}
\centering
\resizebox{0.45\textwidth}{!}{%
\includegraphics{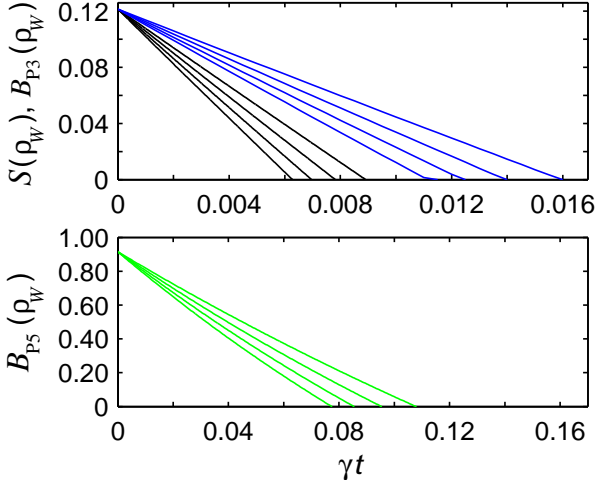}}
\caption{(Color online) Bell-nonlocality violation of $\rho_W(t)$.
Here the black, blue and green curves show respectively dynamics of
$S(\rho_W)$, $B_{\rm P3}(\rho_W)$ and $B_{\rm P5}(\rho_W)$. For
every line color, the curves from right to left correspond to the
cases of $\bar{n}=0$, $\bar{n}=0.1$, $\bar{n}=0.2$ and
$\bar{n}=0.3$.} \label{fig:5}
\end{figure}

Now we discuss Bell-nonlocal behaviors of $\rho_W(t)$. The
measurement operators associated with the first qubit of the initial
{\it W} state are $M_A\equiv\sigma_z$ and $M'_A\equiv\sigma_x$
\cite{Jaeger}, while for the second and the third qubits they are
still defined as those in Eq. (10). Then the expectation value of
the Svetlichny operator $\mathcal {S}$ can be readily obtained as
\begin{eqnarray}
 \langle \mathcal {S}\rangle_{\rho_{W}(t)}&=&(2\rho_W^{11+44+66+77+46+47+67-23-25-35}-1)
                                           \nonumber\\&&
                                           \times(\sin\theta_{BC}+\cos\theta_{BC}),
\end{eqnarray}

For the zero temperature reservoir (i.e., $\bar{n}=0$), the maximum
value of $\langle \mathcal {S}\rangle_{\rho_{W}(t)}$ simplifies to
$|\langle \mathcal
{S}\rangle_{\rho_{W}(t)}|=(4+5\sqrt{2})p^4-(2+9\sqrt{2}/2)p^2+\sqrt{2}$,
which decays with time and becomes smaller than 4 in a finite
timescale $\gamma\tau_{\rm
S}=\ln[(16+20\sqrt{2})/(4+9\sqrt{2}+\sqrt{274+328\sqrt{2}})]\simeq0.00891$.
For the nonzero temperature reservoirs (i.e., $\bar{n}\neq 0$),
since the density matrix elements of $\rho_W(t)$ are so complicated,
we define $S(\rho_W)={\rm max}\{|\langle\mathcal {S}\rangle_{\rho_W
(t)}|-4,0\}$ and present its dynamical behaviors numerically in Fig.
5 as black curves. Clearly, $S(\rho_W)$ decays monotonically with
increasing $\gamma t$ and becomes zero in a finite time $\tau_{\rm
S}$, after which the genuinely tripartite Bell-nonlocal correlation
disappears. $\tau_{\rm S}$ decreases with increasing temperature,
and as can be found obviously from Fig. 5, its magnitude is very
small $(\gamma\tau_{\rm S}<0.00891)$, i.e., the genuinely tripartite
Bell nonlocality for the initial {\it W} state preparation is
fragile and the thermal reservoir destroys it in a very short
timescale.

The expectation values for the five WWZB operators $\mathcal
{B}_{\rm PI}$ $({\rm I}=1,2,3,4,5)$ can also be obtained in terms of
the density matrix elements of $\rho_W(t)$ as
\begin{eqnarray}
 \langle\mathcal {B}_{\rm P1}\rangle_{\rho_{W}(t)}&=&(4\rho_W^{11+44+66+77}-2)\cos\theta_{B}\cos\theta_{C}
                                                      \nonumber\\&&
                                                      +4\rho_W^{23-47}\sin\theta_{B}\sin\theta_{C},\nonumber\\
 \langle\mathcal {B}_{\rm P2}\rangle_{\rho_{W}(t)}&=&\rho_W^{25+35-46-47}\cos\theta_{BC}
                                                     \nonumber\\&&
                                                     +\rho_W^{25-35+46-47}\sin(\theta_{B}-\theta_{C})
                                                     \nonumber\\&&
                                                     +(\rho_W^{23-67-11-44-66-77}+1/2)
                                                     \nonumber\\&&
                                                     \times(\cos\theta_{BC}-\sin\theta_{BC}),\nonumber\\
 \langle\mathcal {B}_{\rm P3}\rangle_{\rho_{W}(t)}&=&(2\rho_W^{11+44+66+77-35+46}-1)x_{+}
                                                     \nonumber\\&&
                                                     +4\rho_W^{23-47}x_{-},\nonumber\\
 \langle\mathcal {B}_{\rm P4}\rangle_{\rho_{W}(t)}&=&(2\rho_W^{11+44+66+77}-1)y_{+}+2\rho_W^{35-46}y_{-}
                                                     \nonumber\\&&
                                                     +4\rho_W^{23-47}\sin\theta_{B}\sin\theta_{C},\nonumber\\
 \langle\mathcal {B}_{\rm P5}\rangle_{\rho_{W}(t)}&=&(2\rho_W^{11+44+66+77+46+47+67-23-25-35}-1)
                                                     \nonumber\\&&
                                                     \times\sin\theta_{BC}.
\end{eqnarray}
where $x_{+}=\sqrt{2}\sin(\theta_{B}+\pi/4)\cos\theta_{C}$,
$x_{-}=\sqrt{2}\sin(\theta_{B}-\pi/4)\sin\theta_{C}$,
$y_{\pm}=\sqrt{2}\sin(\theta_{C}\pm\pi/4)\cos\theta_{B}$, and in
deriving the above equations, we have used the hermiticity condition
of the density matrix.

We define $B_{\rm PI}(\rho_W)={\rm max}\{|\langle\mathcal {B}_{\rm
PI}\rangle_{\rho_W(t)}|-2,0\}$ $({\rm I}=1,2,3,4,5)$ to evaluate the
extent to which the WWZB-type inequalities are violated. Still due
to the fact that the elements of $\rho_W(t)$ are so complicated, it
is difficult to express $B_{\rm PI}(\rho_W)$ compactly with respect
to the parameters $\gamma t$ and $\bar{n}$. Thus we resort to
numerical calculations. The results show that the inequalities of
forms ${\rm P1}$, ${\rm P2}$ and ${\rm P4}$ are satisfied in the
whole time region (i.e., $B_{\rm PI}(\rho_W)\equiv 0$ for ${\rm
I}=1,2,4$). For the remaining two classes of forms ${\rm P3}$ and
${\rm P5}$, examples of decays of $B_{\rm P3}(\rho_W)$ and $B_{\rm
P5}(\rho_W)$ with different $\bar{n}$ are presented graphically in
Fig. 5 as blue and green curves, respectively. It is obvious that
they decay monotonously with increasing $\gamma t$ and become zero
after finite times $\tau_{B_{\rm P3}}$ and $\tau_{B_{\rm P5}}$ (when
$\bar{n}=0$ we have $\gamma\tau_{B_{\rm
P5}}=\ln[(20+8\sqrt{2})/(9+2\sqrt{2}+\sqrt{169+68\sqrt{2}})]\simeq0.10785$),
both of which decrease with increasing $\bar{n}$. The time
$\tau_{B_{\rm P5}}$ demonstrates the time of sudden death of all
species of Bell-nonlocal correlations for $\rho_W(t)$ under the
influence of thermal noise. Particularly, by comparing the present
results with those of $\rho_{\rm GHZ}(t)$, one can note that
$\tau_{B_{\rm P5}}$ here is very small, i.e., the Bell nonlocality
for $\rho_W(t)$ is very fragile compared with that of $\rho_{\rm
GHZ}(t)$.

Furthermore, one can note that the death time for Bell nonlocality
is much earlier than that for the tripartite negativity or the
critical time after which the teleportation fidelity $F_{\rm
av}(\rho_{W})<2/3$. This reveals several common features. First, as
mentioned before, the positivity of $N(\rho_W)$ ensures the
entanglement of  the state $\rho_{W}(t)$, thus there exist time
regions during which $\rho_W(t)$ possesses local correlations even
in correspondence to high values of entanglement. Second, only
partial of the tripartite entangled states are useful for
nonclassical teleportation. But all the Bell-nonlocal states yield
$F_{\rm av}(\rho_{W})>2/3$.

Finally, we would like to emphasize that even when one or two of the
participating qubits can be well preserved, e.g., $\gamma_A=0$ or
$\gamma_A=\gamma_B=0$, the tripartite negativity and the Bell
nonlocality still experience sudden death, although in a
comparatively longer time. This shows that the effects of thermal
reservoir on entanglement and coherence of a qubit is indeed very
different. Since they change in the similar manner as that of
$\gamma_A=\gamma_B=\gamma_C=\gamma$ and no other new results can be
drawn, we won't have any more discussions about them here.

\section{Summary}
In summary, we have investigated behaviors of disentanglement and
Bell-nonlocality violation for various decaying states and analyzed
their relations with capacity of these states when being used as
quantum channels for teleportation. Our system consists of three
qubits prepared initially in the GHZ or {\it W} class state and
coupled to a thermal reservoir, under the influence of which
irreversible coherence decay will be unavoidable. Depending on the
temperature of the reservoir, the tripartite negativity can reach
value equal to zero asymptotically (if $\bar{n}=0$) or at a finite
time (if $\bar{n}\neq 0$). But the sudden death of tripartite
correlations associated with the Svetlichny inequality and nonlocal
correlations associated with the WWZB inequalities are irreversible
in the whole temperature regions. Moreover, the tripartite
negativity and Bell nonlocality for $\rho_{\rm GHZ}(t)$ are more
robust than that for $\rho_W(t)$ in the sense that they survive in a
significantly longer times under the influence of thermal reservoir.
Particularly, $\rho_{\rm GHZ}(t)$ gives a wider time region for
nonclassical teleportation. Finally, by comparing the survival time
for Bell nonlocality with the time region during which $F_{\rm
av}(\rho_{\rm \Pi})>2/3$ ($\Pi={\rm GHZ}$ or $W$), we showed that
all the Bell-nonlocal states considered in this work can be used for
quantum teleportation, while there also exist a family of entangled
mixed states which do not violate any multipartite Bell-type
inequalities, but still yield nonclassical teleportation fidelity.
\\

\begin{center}
\textbf{ACKNOWLEDGMENTS}
\end{center}

This work was supported by the NSF of Shaanxi Province under Grant
Nos. 2010JM1011 and 2009JQ8006, the Scientific Research Program of
Education Department of Shaanxi Provincial Government under Grant
No. 2010JK843, and the Youth Foundation of XUPT under Grant No.
ZL2010-32.

\newcommand{\PRL}{Phys. Rev. Lett. }
\newcommand{\PRA}{Phys. Rev. A }
\newcommand{\JPA}{J. Phys. A }
%

%

\end{document}